\documentclass[final,5p,times,number,natbib]{elsarticle}%FINAL

\usepackage{amsmath}
\usepackage{amssymb}
\usepackage{amsfonts}
\usepackage{color}
\usepackage[normalem]{ulem}
\usepackage{ifpdf}\ifpdf % si pdftex
\usepackage[colorlinks,linktocpage,pdftex,hyperfigures]{hyperref}
\makeatletter
\providecommand{\doi}[1]{%
  \begingroup
    \let\bibinfo\@secondoftwo
    \urlstyle{rm}%
    \href{http://dx.doi.org/#1}{%
      doi:\discretionary{}{}{}%
      \nolinkurl{#1}%
    }%
  \endgroup
}
\makeatother

\fi\usepackage{lineno}
\biboptions{sort&compress}

\journal{Journal of Colloid and Interface Science}

\begin{document}

\begin{frontmatter}

\title{Can hydrodynamic contact line paradox be solved by evaporation-condensation?}

\author[fast,am]{V. Jane\v{c}ek}
\author[fast,upmc]{F. Doumenc\corref{cor1}}
\cortext[cor1]{Corresponding author} \ead{doumenc@fast.u-psud.fr}
\author[fast]{B. Guerrier}
%\ead{guerrier@fast.u-psud.fr}
\author[SPEC]{V. S. Nikolayev}
%\ead{Vadim.Nikolayev@cea.fr}
\address[fast]{University Paris-Sud, CNRS, Lab FAST, Bat 502, Campus Universitaire, Orsay 91405, France}
\address[am]{Present address: ArcelorMittal, Voie Romaine, BP 30320,  Maizi\`eres-l\`es-Metz, 57283,   France}
\address[upmc]{ Sorbonne Universit\'es, UPMC Univ Paris 06, UFR 919, 75005, Paris, France  }
\address[SPEC]{Service de Physique de l'Etat Condens\'e, CNRS UMR 3680, IRAMIS/DSM/CEA Saclay, 91191 Gif-sur-Yvette, France}

\begin{abstract}
We investigate a possibility to regularize the hydrodynamic contact line singularity in the configuration of partial wetting (liquid wedge on a solid substrate) via evaporation-condensation,
when an inert gas is present in the atmosphere above the liquid.
The no-slip condition is imposed at the solid-liquid interface and the system is assumed to be isothermal. The mass exchange dynamics is controlled by vapor diffusion in the inert gas and interfacial kinetic resistance. The coupling between the liquid meniscus curvature and mass exchange is provided by the Kelvin effect.
The atmosphere is saturated and the substrate moves at a steady velocity with respect to the liquid wedge.
A multi-scale analysis is performed. The liquid dynamics description in the phase-change-controlled microregion and visco-capillary intermediate region is based on the lubrication equations. The vapor diffusion is considered in the gas phase.
It is shown that from the mathematical point of view, the phase exchange relieves the contact line singularity. The liquid mass is conserved: evaporation existing on a part of the meniscus and condensation occurring over another part compensate exactly each other. However, numerical estimations carried out for three common fluids (ethanol, water and glycerol) 
at the ambient conditions show that the characteristic length scales are tiny. 
\end{abstract}


\end{frontmatter}

\section{Introduction}

Since the seminal article by \citet{Huh}, it is well known that the standard hydrodynamics fails in describing the motion of the triple liquid-gas-solid contact line in a configuration of partial wetting.
Their hydrodynamic model based on classical hydrodynamics with the no-slip condition at the solid-liquid interface and the imposed to be straight liquid-gas surface predicts
infinitely large viscous dissipation. If the normal stress balance is considered at the free surface, such a problem has no solution at all \citep{Pismen11}. As an immediate consequence, a droplet cannot slide over an inclined plate, or a solid cannot be immersed into a liquid.

Despite the fact that this paradox is known for decades, it is still a subject of intense debate (see for instance \cite{Velarde11}).

Contact line motion is in fact a multi-scale problem, and  microscopic effects must be considered in the vicinity of the contact line to solve the above-mentioned paradox (see \cite{Bonn09,Snoeijer13} for reviews).
One can make a distinction between approaches for which the dissipation is located at the contact line itself, from models where dissipation is assumed to be of viscous origin, inside the liquid.
In the former class of models, referred as molecular kinetic theory, the contact line motion is driven by jumps of molecules close to
the contact line \citep{BH1}. In the latter approach, based on hydrodynamics, some microscopic features are to be included.
\citet{Hocking83,Anderson,PF10} solved such a problem by incorporating the hydrodynamic slip.
In the complete wetting case, the van der Waals forces cause a thin adsorbed film over the substrate, which relieves the singularity.
For such a case, \citet{Moosman80, DasGupta93, Morris01, Rednikov11} considered the pure vapor atmosphere and the substrate superheating. \citet{Poulard05, Pham10, Eggers10, Doumenc11, Morris14} investigated the diffusion-limited evaporation, when an inert gas is present in the under-saturated atmosphere. Up to now, the case of partial wetting and diffusion-controlled phase change received less attention.
\citet{Berteloot08} proposed an approximate solution for an infinite liquid wedge on a solid substrate using the expression of the evaporation flux given by Deegan \textit{et al} \cite{Deegan00}.
The singularity is avoided by assuming a finite liquid height at a microscopic cut-off distance, imposed \textit{a priori}.

\citet{Wayner93} suggested that the contact line could move by condensation and evaporation while the liquid mass is conserved. During the advancing motion, for instance,
the condensation may occur to the liquid meniscus near the contact line while the compensating evaporation occurs at another portion of the meniscus. Such an approach seemed very attractive \cite{Pomeau,JFM02} since it could provide a model with no singularity
although completely macroscopic, avoiding microscopic ingredients such as slip length or intermolecular interactions.
Rigorous demonstrations of the fact that change of phase regularizes the contact line singularity has been done recently by two independent groups \cite{EuLet12,Rednikov13,PRE13movingCL},
for the configuration of a liquid surrounded by its pure vapor. In this configuration, evaporation or condensation rate is controlled by the heat and mass exchange phenomena in the liquid. Such a situation occurs e.g. for bubbles in boiling. 
The Kelvin effect has proved to be very important because it provided a coupling between the liquid meniscus shape and mass exchange.
In the present work, we explore a possibility of relaxation of the contact line singularity by the phase change in the contact line vicinity in a common situation where a volatile liquid droplet is surrounded by an atmosphere of other gases like air.
This case is more challenging than the case of the pure vapor, because the evaporation or condensation rate is controlled by the vapor diffusion in the gas, which results in non-local evaporation or condensation fluxes \citep{Eggers10}.

\begin{figure}[tb]
\centering\includegraphics[width=0.4\textwidth,clip]{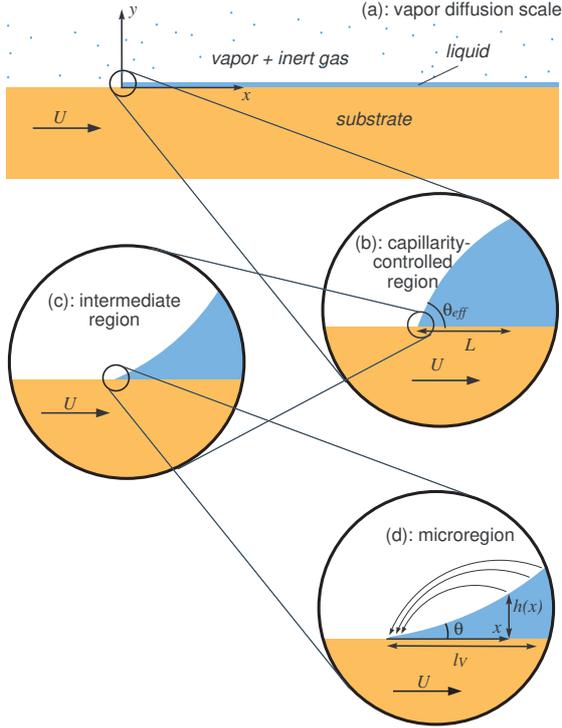}
\caption{Hierarchy of scales considered in the article and geometries for the vapor diffusion and hydrodynamic problems. The curved arrows in the microregion (d) show the vapor diffusion fluxes associated with evaporation-condensation.} \label{geom}
\end{figure}

The following physical phenomena need to be accounted for in such a problem.
\begin{itemize}
	\item The concentrational Kelvin effect, i.e. a dependence of the saturation vapor concentration on the meniscus curvature. 
This effect is expected to be important in a small region of the liquid meniscus very close to the contact line, that we call microregion (Fig. \ref{geom}d). In this region, high meniscus curvature is associated to the strong evaporation or condensation. The microregion size is expected to be below 10-100 nm.
	\item A region of mm scale, where the surface curvature is controlled by the surface tension, and (depending on the concrete macroscopic meniscus shape) gravity or inertia (Fig. \ref{geom}b). The viscous stresses associated with the contact line motion and phase change are negligible here.
	\item A region of intermediate scale (Fig. \ref{geom}c), where both capillary forces and viscous stresses are important.	This region is known to be described by the Cox-Voinov relation \citep{Voinov,Cox86}
\begin{equation}
	\label{CoxVoinov}
	h'(x)^3 = \theta_V^3 + 9 \/ Ca \ln(x/\ell_V),
\end{equation}
with $h'(x)$ the liquid slope at a distance $x$ from the contact line and $Ca=\mu U / \sigma$ the capillary number
($\mu$ is the liquid viscosity, $\sigma$ the surface tension and $U$ the contact line velocity, assumed to be positive for the advancing contact line). It is a solution of Stokes equations in lubrication approximation that satisfies the boundary condition of vanishing curvature at large $x$. Note that large at the intermediate scale $x$ remains small at the macroscopic scale associated with the macroscopic radius $L$ of meniscus curvature (defined e.g. by the drop size when controlled by capillarity). Similarly, the curvature $L^{-1}$ can be considered as negligible with respect to curvatures induced by strong viscous stresses in the intermediate region. Eq. \eqref{CoxVoinov} is valid for small capillary numbers, below the Landau-Levich transition for the receding contact line \citep{Snoeijer13}. $\ell_V$ is a length of the order of the microregion size and is called the Voinov length while $\theta_V$ is the Voinov angle. The Cox-Voinov relation provides a good description of the intermediate region because of the strong scale separation between the capillarity controlled region and microregion. In contact line motion models, the Voinov length and angle can be obtained by the asymptotic matching to the microregion, while the asymptotic matching to the capillarity controlled region provides the following relation for the effective contact angle $\theta_{eff}$ (cf. Fig. \ref{geom}b),
\begin{equation}
        \label{CoxVoinovEff}
	\theta_{eff}^3 = \theta_V^3 + 9 \/ Ca \ln(L/\ell_V).
\end{equation}
The $L$ value depends on the concrete macroscopic meniscus shape \citep{Snoeijer13}. Since we are interested in the relaxation of the contact line singularity, the capillarity-controlled region is not considered here,
and the liquid meniscus is assumed to be a liquid wedge in both the intermediate region and microregion.
	\item Because of the long range of the concentration field controlled by vapor diffusion in the air, one needs to consider one more scale much larger than that of the liquid meniscus. In the following, we assume that at this scale the liquid meniscus is a semi-infinite ($x\in[0,\infty]$) layer of the negligibly small height that covers the solid substrate (Fig. \ref{geom}a).
\end{itemize}

\section {Problem statement} \label{wms}

The problem to be considered is a liquid wedge posed on a flat and homogeneous substrate moving at constant velocity $U$, in a situation of partial wetting.
The atmosphere surrounding the substrate and the liquid consists of an inert gas saturated with the vapor of the liquid, cf. Fig. \ref{geom}a
(an instance of such an atmosphere is wet air at atmospheric pressure, room temperature and relative humidity of 100\%).
The problem is assumed to be isothermal, which may be justified when the substrate is a good thermal conductor.
The vapor concentration is imposed at an infinite (at the scale of Fig. \ref{geom}a) distance from the substrate, and corresponds to the saturated vapor pressure.
Therefore, the system is at equilibrium when $U=0$.
When the substrate moves, viscous pressure drop induces free surface bending and a non zero curvature (Figs. \ref{geom}c,d).
Because of the Kelvin effect, this deformation results in a change of the equilibrium vapor pressure above the gas/liquid interface, leading to evaporation and condensation (Fig. \ref{geom}d). The model considers the vapor diffusion in the gas phase, as well as the kinetic interfacial resistance.

\subsection {Governing equations} \label{eqConfA}

\subsubsection {Liquid phase}

Within the lubrication approximation (small contact angles) the governing equation in the contact line reference is
\begin{eqnarray}\label{Eq_hydro}
\frac{d}{dx}\left(\frac{h^3}{3\mu}\frac{d(\sigma \kappa)}{d x}\right)&=&-U\frac{dh}{dx}-\frac{j}{\rho}, \label{Young_laplace_eq1}
\end{eqnarray}
where $h$ is the liquid height, $\mu$ the liquid viscosity, $\sigma$ the surface tension, $\kappa$ the curvature ($\kappa \simeq d^2h/dx^2$ in the framework of small slope approximation),  $ U $ the substrate velocity  ($U>0$ for an advancing contact line),  $\rho$ the liquid density, and $j$ the interfacial mass flux ($j>0$ for evaporation).

Note that equation \eqref{Young_laplace_eq1} can be rewritten as ${dq}/{dx}=-j/\rho$,
where $q(x)$ is the fluid volume flux in the liquid layer vertical section.

Boundary conditions for the liquid phase are the following:
\begin{itemize}
\item For $x=0$ (contact line):
\begin{equation}\label{BC_liquid1}
h = 0,\,\frac{dh}{dx}=\theta,
\end{equation}
where $\theta $ is the equilibrium contact angle imposed by intermolecular interactions at micro scale.
\end{itemize}
\begin{itemize}
\item
For  $x\to \infty$:
\begin{equation}\label{BC_liquid_31}
\frac{d^2h}{dx^2} = 0.
\end{equation}
\end{itemize}
In addition, since we are interested in  regularization through evaporation/condensation, we look for regular solutions. In other words, the mass flux $j(x)$, the curvature $\kappa (x)$ and $d\kappa (x)/dx$ (and thus the pressure gradient) are assumed to be finite  at $x=0$.

\subsubsection{Gas phase and mass flux}
 In the gas phase, the equation for vapor diffusion reads
\begin{equation}\label{Laplace_eq1}
\frac{\partial ^2 c}{\partial x^2}+\frac{\partial ^2 c}{\partial y^2}=0,
\end{equation}
where $c(x,y)$ is the vapor concentration.
In the framework of the small wedge slope approximation, the liquid gas interface seen from the large scale of the gas atmosphere is assumed to coincide with the line $y=0, x\geq 0$, as shown in Fig. \ref{geom}a.
The boundary conditions in partial wetting configuration are:
\begin{itemize}
\item For $x\to\pm\infty$ or $y\to \infty$:
\begin{equation}
c(x,y)=c_{eq},
\label{bcinf}
\end{equation}
where $c_{eq}$ is the vapor concentration  at thermodynamic equilibrium for a flat (with $\kappa=0$) liquid-gas interface.
\item For $y=0$ and $x<0$:
\begin{equation}
\frac{\partial c}{\partial y}=0.
\end{equation}
\item For $y=0$ and $x\geq 0$:
\begin{equation}
\label{interface_resistance1}
  j = \frac{c_{i,eq}-c_{i}}{R_i} = \left.-D_g\frac{\partial c }{\partial y}\right|_{y=0},
\end{equation}
where $c_i =c(x,y=0) $ is the vapor concentration at the liquid-gas interface and $c_{i,eq}$ is the equilibrium interfacial vapor concentration that depends on the interface curvature $\kappa $ through the concentrational Kelvin equation that reads
\begin{equation}\label{concentration_linear1}
  c_{i,eq}-c_{eq}=-\frac{Mc_{eq}}{\rho R_g T}\sigma\kappa\end{equation}
in its linearized version \citep{Eggers10}. $D_g$ is the vapor diffusion coefficient in the gas phase and  $R_i $ is the kinetic resistance given by the Hertz-Knudsen relation \citep{Barnes86,Carey},
\begin{equation} \label{HertzKnud}
R_i^{-1}=\frac{2f}{2-f}\sqrt{\frac{R_g T}{2\pi M}},
\end{equation}
where $f$ is the accommodation factor close to unity, $R_g$ is the ideal gas constant, $T$ is the  temperature and $M$ is the molar mass.
\end{itemize}

\subsubsection{Kelvin length and dimensionless equations}\label{1scales}

Let us determine the Kelvin length $\ell $, a characteristic size of a region dominated by the Kelvin effect, with the following scaling analysis.
The dimensionless abscissa is $\tilde{x}=x/\ell $ and, accordingly to the wedge geometry, the liquid height scales as $H=h/(\theta\ell )$.
The dimensionless curvature is hence $\mathcal K=\ell \kappa/\theta$.
Introducing these expressions in equation (\ref{Young_laplace_eq1}) brings out  the modified capillary number $\delta= 3 Ca/\theta^3 $ and a characteristic scale $J=\theta^4\sigma \rho/(3\mu)$ for the mass flux ($\tilde{j} = j/J$):
 \begin{equation} \label{scaling0}
\frac{d}{d\tilde{x}}\left(H^3\frac{d \mathcal K}{d \tilde{x}}\right)= -\delta \frac{dH}{d\tilde{x}}- \tilde{j}.
\end{equation}

The vapor concentration above the liquid/gas interface, $c_i$, is related to the film curvature $\kappa $ via the Kelvin effect.
Therefore, $c_i$ also varies in $x$-direction over a length of the order of $\ell $.
Moreover,  due to the Laplace equation in the gas phase, the length scales for $x$ and $y$ directions should be the same, thus $\tilde{y}=y/\ell $.
Concentration deviation is reduced as $\tilde{c}=(c-c_{eq})/C$.
By using equation (\ref{concentration_linear1}), one gets the $C$ value
\begin{equation}\label{curv}
 C=\frac{M \sigma \theta c_{eq}}{\rho R_g T \ell} \;\;\; {\rm and } \;\;\; \tilde{c}_{i,eq} = - \mathcal K.
\end{equation}
One notes that $C$ and $J$ are the typical scales of the concentration deviation from equilibrium and mass flux caused by the Kelvin effect, so that both $\tilde{j}$ and $\mathcal K$ are considered to be of the order 1 for the scaling analysis purposes.

Scaling analysis of equation (\ref{interface_resistance1}) reads
$$
 j \sim \frac{c_{i,eq}-c_{i}}{R_i} \sim  D_g \frac{c_i - c_{eq}}{\ell }.
$$
By using equation (\ref{curv}), one gets in scaled variables
\begin{equation} \label{j1}
	\left ( \frac{J \mathcal R \ell}{C D_g} \right )  \tilde j \sim (\mathcal K + \tilde{ c}_i) \sim  \mathcal R \tilde {c}_i ,
\end{equation}
where $\mathcal{R}=R_i D_g/\ell $ is the dimensionless interfacial resistance.
To get the scaling of $\tilde j$, a balance of three terms is to be discussed,
\begin{equation} \label{3T}
 \left[  \mathcal K\; ; \;   \tilde {c}_i \; ; \; \mathcal R\tilde {c}_i \right].
\end{equation}

One can distinguish two limiting cases.
\begin{itemize}
\item Case 1, $\mathcal R \gg 1$:  the kinetic resistance $R_i $ dominates the vapor diffusion resistance $\ell / D_g$ to the mass transfer at the interface. The second term in the set (\ref{3T}) is negligible with respect to the third. The balance between the curvature (first term related to the Kelvin effect) and the third term reads $\tilde {c}_i\sim\mathcal K/\mathcal R$. From equation~(\ref{j1}) one thus has
\begin{equation}\label{scale}
\tilde j\sim\frac{C D_g\mathcal K}{J\ell\mathcal R}.
\end{equation}
Since both $\mathcal K$ and $\tilde j$ are of the order unity, one can obtain from eq. \eqref{scale} the characteristic length $\ell$ that we call $\ell_R$ for this case:$$\ell_R  = \frac{C D_g} {J \mathcal R}= \frac {3 \mu  M c_{eq}}{\theta^3 \rho ^2R_g T R_i }.$$
\item Case 2, $\mathcal R \ll 1$: this is the opposite case corresponding to the negligible kinetic resistance ($R_i \ll \ell / D_g$), the second term in the set (\ref{3T}) has to balance the curvature:
	$$ \tilde {c}_i \sim \mathcal K  \;\;\; {\rm thus} \;\;\;  \tilde j  \sim \frac{C D_g  \mathcal K}{J \ell } ~ ~ ~ \rm{from~equation~(\ref{j1})}, $$
which leads to the characteristic length $$\ell_D = \frac{C D_g}{J }  = \frac{1}{\rho} \sqrt{ \frac{3 \mu  M c_{eq} D_g}{\theta^3 R_g T }}.$$
\end{itemize}
One can see that $\ell_D$ is the Kelvin length for the diffusion regime,
while $ \ell_R= \ell_D /\mathcal R $ is the Kelvin length for the kinetic regime.
Throughout the rest of the paper, we choose $\ell = \ell_D$ to make the equations dimensionless.
Numerical estimations of the relevant scales are given in section \ref{usliq} for three common fluids for the ambient conditions.

The dimensionless lubrication equation  (\ref{Eq_hydro}) reads
\begin{equation}\label{Eq_hydro_nond}
\frac{d}{d\tilde{x}}\left(H^3\frac{d \mathcal K}{d \tilde{x}}\right)=-\delta\frac{dH}{d\tilde{x}}-\tilde{j},
\end{equation}
with the following boundary conditions:
\begin{itemize}
\item For $\tilde{x}=0:\quad H=0,\quad H'=1,$
\item For $\tilde{x}\to \infty: \quad \mathcal K = H''=0$.
\end{itemize}

The dimensionless diffusion equation in the gas phase is
\begin{equation}\label{Laplace_eq_nond}
\Delta\tilde{c}\equiv\frac{\partial ^2 \tilde{c}}{\partial \tilde{x}^2}+\frac{\partial ^2 \tilde{c}}{\partial \tilde{y}^2}=0,
\end{equation}
with the following boundary conditions:
\begin{itemize}
\item For $\tilde{x}\to\pm\infty$ or $\tilde{y}\to \infty$:
\begin{equation} \label{CLinf}
\tilde{c}=0,
\end{equation}

\item For $\tilde{y}=0$ and $\tilde{x}<0$:
\begin{equation} \label{CLsub}
\frac{\partial \tilde{c}}{\partial\tilde{y}}=0,
\end{equation}

\item For $\tilde{y}=0$ and $\tilde{x}\geq0$:
\begin{equation}
\tilde{j}=-\frac{\mathcal{K}+\tilde{c}}{\mathcal R}= -\frac{\partial \tilde{c}}{\partial \tilde{y}}.\label{mass_difusion_flux_nond}
\end{equation}
\end{itemize}

\subsection{Mass conservation issue}\label{conserv}

Before solving the problem, let us consider the mass conservation issue in the gas domain $\mathcal D$, cf. the upper half plane in Fig. \ref{geom}a.
Let us apply the divergence (Gauss) theorem to $\nabla \tilde c$,
\begin{equation}\label{dt}
\int_\mathcal D \Delta \tilde c(\vec{r})d\vec{r}=\oint_\mathcal L \frac{\partial \tilde c}{\partial\vec{n}} dl,\end{equation}
where $\mathcal L$ is the boundary of $\mathcal D$, and ${\partial \tilde c}/{\partial\vec{n}} \equiv \vec{n}\cdot\nabla\tilde c$.
The contour $\mathcal L$ consists of the $x$-axis and a half-circle $\mathcal C$ of infinite radius in the upper half-plane.
From the vapor diffusion equation \eqref{Laplace_eq_nond}, one obtains the mass conservation in $\mathcal D$:
\begin{equation}\label{mconsL}
\oint_\mathcal L \frac{\partial \tilde c}{\partial\vec{n}} dl=0.
\end{equation}
We show below that the manifestation of the mass conservation is
\begin{equation}\label{mconsx}
\int_0^\infty \tilde j(\tilde x) d\tilde x=0,
\end{equation}
i.e. the flux through $\mathcal C$ is zero (cf. Eq.(\ref{CLsub})).

Let $Q(r)$ be the flux through $\mathcal C_r$, a half circle of radius $r$ that tends to $\mathcal C$ when $r\to\infty$,
\begin{equation} \label{Qrint}
Q(r)= \int_{\mathcal C_r} \frac{\partial \tilde c}{\partial\vec{n}} dl=\int_0^\pi \frac {\partial \tilde c}{\partial r}\; r \; d\phi,
\end{equation}
with $\phi $ the polar angle.
After dividing both members of equation (\ref{Qrint}) by $r$, and integrating over $r$ from some arbitrary value $r_0$ to $\infty$, one gets because of the condition \eqref{CLinf}
\begin{equation} \label{Qrint2}
\int_{r_0}^\infty \frac{Q(r)}{r} \; dr= -\int_0^\pi \tilde c(r_0,\phi ) \; d\phi,
\end{equation}
where the right hand side is finite.
The integral in the left hand side can only be convergent if $Q(r) \to 0$ when $r \to \infty$ which proves equation (\ref{mconsx}). It means that the overall mass flux transferred to the gas environment is zero, which is consistent with the assumption of thermodynamic equilibrium at $r \to \infty$.
This conclusion is quite important. It means that the evaporation and condensation fluxes
compensate exactly each other during the contact line motion so that the liquid mass is conserved. Since the exact compensation is due to the vapor diffusion, it may be violated in its absence (interfacial resistance-controlled phase change).

The vanishing at infinity flux $Q(r)$ implies
that $\partial \tilde c / \partial r$ goes to zero faster than $r^{-1}$ when $r \to \infty$, cf. equation (\ref{Qrint}). This fact is used in \ref{appA} to derive the governing equation.

\subsection{First order approximation}
The  variables are expanded  in a regular perturbation series in  the modified capillary number $\delta$. At the zero order corresponding to  motionless substrate and thermodynamic equilibrium, obviously $\mathcal K_0=0$, $H_0=\tilde{x}$, $\tilde{j}_0=0$ and $\tilde{c}_0 =0$, thus
\begin{eqnarray}
\mathcal K &=& \delta \mathcal K_1+\cdots, \label{lubper}\nonumber\\
H &=& \tilde x +\delta H_1+\cdots, \label{lubper2}\\
\nonumber
\tilde{j} &=& \delta \tilde{j}_1+\cdots, \label{lubper3}\\
\nonumber
\tilde{c} &=& \delta \tilde{c}_1+\cdots.
\label{lubper4}
\end{eqnarray}
The lubrication equation (\ref{Eq_hydro_nond}) at the first order  reads
\begin{equation}\label{Eq_hydro_nond_first_order}
\frac{d}{d\tilde{x} }\left(\tilde{x}^3 \frac{d {\mathcal K_1}}{d\tilde{x} }\right)= -1-\tilde{j}_1,
\end{equation}
with the following boundary conditions:
\begin{itemize}
\item For $\tilde{x}=0: H_1=0, H'_1=0,$
\item For $\tilde{x}\to \infty: \mathcal K_1 =  H''_1=0.$
\end{itemize}

The first order problem for the diffusion in the gas phase coincides with that for $\tilde{c}$, eqs. (\ref{Laplace_eq_nond}-\ref{mass_difusion_flux_nond}).

\section{Asymptotic solution in the intermediate region \label{CVsection}}

As mentioned in the introduction, the intermediate region (Fig. \ref{geom}c) is characterized by a balance between viscous stress and capillary pressure.
From the scaling analysis in section \ref{1scales}, one infers that the cross-over between the microregion (dominated by Kelvin effect) and the intermediate region should occur at
$\tilde x \sim 1$ in the diffusive regime ($\mathcal R \ll 1$) and $\tilde x \sim 1 / \mathcal R$ in the kinetic regime ($1 \ll \mathcal R$).
Evaporation/condensation fluxes are induced by Kelvin effect only.
Therefore, for $\tilde x \gg \min(1,1/\mathcal R)$, the absence of the Kelvin effect implies $\tilde j_1 =0$ and the problem becomes that of Cox-Voinov.
Equation (\ref{Eq_hydro_nond_first_order}) may be integrated over the intermediate region resulting in \begin{equation}\label{Cox_Voinov_1}
	\tilde q_1(\tilde x)\equiv\tilde{x}^3 \frac{d {\mathcal K_1}}{d\tilde{x} } + \tilde x=\alpha,
\end{equation}
where $\alpha$ is an integration constant.

The overall evaporation and condensation in the microregion compensate each other, see section \ref{conserv}, so that there is no flow at the upper microregion boundary (that corresponds to $\tilde x\to 0$ in the intermediate region). The horizontal flow flux $\tilde q_1( \tilde x)$ must thus vanish in the beginning of intermediate region. This fixes $\alpha=0$ and an integration of equation \eqref{Cox_Voinov_1} results in
\begin{equation} \label{CIR}
\mathcal K_{1}= \frac{1}{\tilde{x}}+ \beta.
\end{equation}
The vanishing curvature at infinity fixes $\beta = 0$. The curvature diverges at small $\tilde{x}$ as expected, since equation \eqref{CIR} is not supposed to be valid in the microregion. One more integration results in the expression
\begin{equation}\label{Cox_Voinovpr}
\frac{dH_{1}}{d\tilde{x}} = \ln\left(\frac{\tilde{x}}{\xi}\right),
\end{equation}
where $\xi$ is an integration constant.
By returning back to the dimensional variables, we get the linearized version of the Cox-Voinov equation (\ref{CoxVoinov})
\begin{equation}\label{Cox_Voinov}
\frac{dh}{dx} = \theta + \frac{3U\mu}{\sigma\theta^2}\ln{\left(\frac{x}{ \ell_V }\right)},
\end{equation}
where  the Voinov length $ \ell_V \equiv \ell_D \xi$ must be deduced from matching to the microregion problem. The Voinov angle $\theta_V$ is equal to $\theta$ because this value of $dh/dx$ corresponds to $U=0$.

\section{Behavior in micro- and intermediate regions} \label{NumSol}
\subsection {Governing equations} \label{reso1}

Assuming that $d \mathcal K_1 / d \tilde{x}$ is finite at $\tilde{x} =0$, integration of equation (\ref{Eq_hydro_nond_first_order}) gives:
\begin{equation}\label{IEq_hydro_nond_first_order}
\tilde{x}^3 \frac{d {\mathcal K_1}}{d\tilde{x} }= -\tilde{x}-\int_0^{\tilde{x}}\tilde{j}_1 d\tilde{x}.
\end{equation}
The first of equalities \eqref{mass_difusion_flux_nond} results in
\begin{equation}\label{curvature1}
\mathcal K_1=-\tilde{c}_{i,1}-\mathcal R\tilde{j}_1.
\end{equation}
The first order concentration in the gas phase at the interface, $\tilde{c}_{i,1}(\tilde x)=\tilde{c}_1(\tilde x,\tilde y=0)$, can be expressed analytically (see \ref{appA}, equations (\ref{convolution_A}) and (\ref{expGA})) as a functional of mass flux,
\begin{align}\label{concentration_at_interface}
&\tilde{c}_{i,1}(\tilde x )=-\frac{1}{\pi}\int_0^{\infty}\ln{|\tilde{x}-\tilde{x}'|}\tilde{j}_1(\tilde{x}')d\tilde{x}'.
\end{align}
Equation (\ref{curvature1}) thus reads
\begin{equation}\label{curvature_convolution}
\mathcal K_1=
\frac{1}{\pi}\int_0^{\infty}\ln{|\tilde{x}-\tilde{x}'|}\tilde{j}_1(\tilde{x}')d\tilde{x}'-
\mathcal R \tilde{j}_1 ,
\end{equation}
and
\begin{equation}\label{Dcurvature_convolution}
\frac {d\mathcal K_1}{d\tilde{x}}=\frac{1}{\pi}\int_0^{\infty}\frac{\tilde{j}_1(\tilde{x}')}{\tilde{x}-\tilde{x}'}
d\tilde{x}'-\mathcal R \frac{d\tilde{j}_1}{d\tilde{x}}.
\end{equation}
Finally, the integro-differential equation governing the mass flux $\tilde{j}_1$ is
\begin{equation}\label{Eq_hydro_nond_first_order_res_T_fin}
\tilde{x}^3\left(\frac{1}{\pi}\int_0^{\infty}\frac{\tilde{j}_1(\tilde{x}')}{\tilde{x}-\tilde{x}'}
d\tilde{x}'-\mathcal R \frac{d\tilde{j}_1}{d\tilde{x}}\right)=\\
-\tilde{x}-\int_0^{\tilde{x}}\tilde{j}_1(\tilde{x}')d\tilde{x}'.
\end{equation}

For arbitrary $\mathcal {R}$, no analytical approach is available and equation (\ref{Eq_hydro_nond_first_order_res_T_fin}) is solved numerically, see \ref{appB}.
Once $\tilde{j}_1$ is known, concentration $\tilde{c}_{i,1}$ and  curvature $\mathcal K_1 $  are obtained through equations
(\ref{concentration_at_interface}) and (\ref{curvature_convolution}) as explained in \ref{appC}. The slope $dH_1 /d\tilde x$ and the height $H_1 $ are then obtained by successive integrations of $\mathcal K_1 $.

\subsection{Purely kinetic regime ($ \mathcal R \to\infty$) } \label{CVsectionK}

For the case dominated by kinetic resistance, it is possible to solve the problem analytically.
When $\mathcal R \gg 1$, the diffusion induced resistance to vapor transfer can be neglected.
Concentration at the interface thus tends to the equilibrium concentration, $\tilde c_{i,1}\to 0$,
and equation (\ref{curvature1}) reduces to $\mathcal K_1=-\mathcal R\tilde{j}_1$.
Equation (\ref{Eq_hydro_nond_first_order})
\begin{equation}\label{Eq_hydro_nond_first_order_res}
\frac{d}{d\tilde{x}}\left[-\tilde{x}^3 \frac{d\tilde{j}_1}{d\tilde{x}}\mathcal R\right]=-1-\tilde{j}_1
\end{equation}
can be then solved analytically,
\begin{equation}\label{solution_}
\tilde{j}_1=-1+\alpha\frac{2 I_2\left(2/\sqrt{\tilde{x}\mathcal{R}}\right)}{\mathcal{R}\tilde{x}}+
\beta\frac{2 K_2\left(2/\sqrt{\tilde{x}\mathcal{R}}\right)}{\mathcal{R}\tilde{x}},
\end{equation}
where $I(\cdot), K(\cdot)$ are the modified Bessel functions of the first and the second order, respectively.
We are looking for a regular solution at $ \tilde{x}=0$, so $\alpha=0$.

At large $\tilde{x}$, we get $$\tilde{j}_1(\tilde{x}\to \infty)\simeq -1+\beta-\frac{\beta}{\mathcal{R}\tilde{x}},$$
with $\beta=1$ to satisfy the flux vanishing at infinity.
The solution is thus
\begin{equation}\label{Solution_flux_only_R}
\tilde{j}_1=-1+
\frac{2 K_2\left(2/\sqrt{\tilde{x}\mathcal{R}}\right)}{\mathcal{R}\tilde{x}}.
\end{equation}
The curvature $\mathcal K_1=-\mathcal R\tilde{j}_1$  then reads
\begin{equation}\label{Solution_flux_only_R_curv}
\mathcal{K}_1=-\frac{2 K_2\left(2/\sqrt{\tilde{x}\mathcal{R}}\right)}{\tilde{x}}+\mathcal{R}.
\end{equation}

\subsection {Curvature and mass flux behavior} \label{CurvFlA}

\begin{figure}[tbh]
\centering
\includegraphics[width=0.4\textwidth,clip]{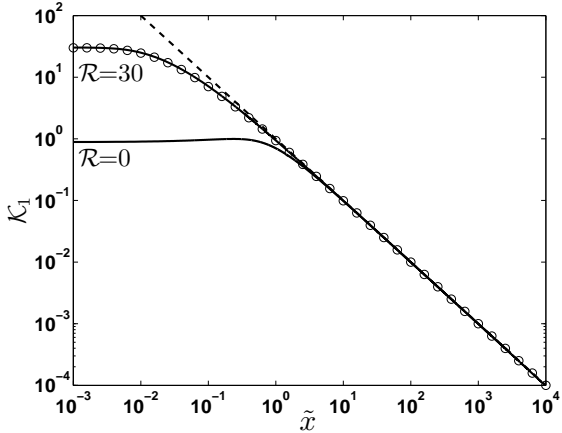}%\\
  \caption{First order curvature term $\mathcal K_1$.
Solid lines: solutions of equation (\ref{Eq_hydro_nond_first_order_res_T_fin}) for $\mathcal {R}=0,30$.
Open circles: purely kinetic regime, equation (\ref{Solution_flux_only_R_curv}) with $\mathcal {R}=30$.
Dashed line: asymptotics (\ref{CIR}). } \label{Curv1}
\end{figure}

\begin{figure}
\centering\includegraphics[width=0.4\textwidth,clip]{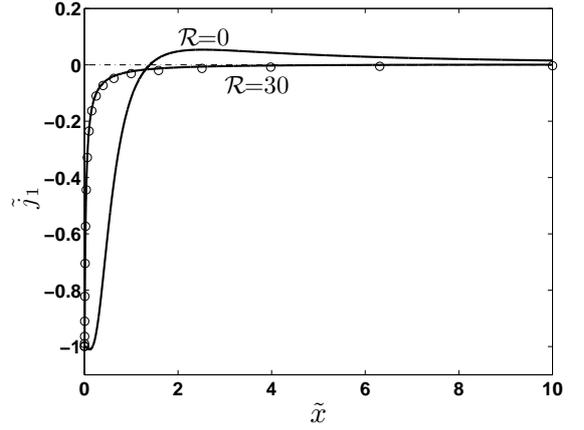}\\
  (a)\\
\includegraphics[width=0.4\textwidth,clip]{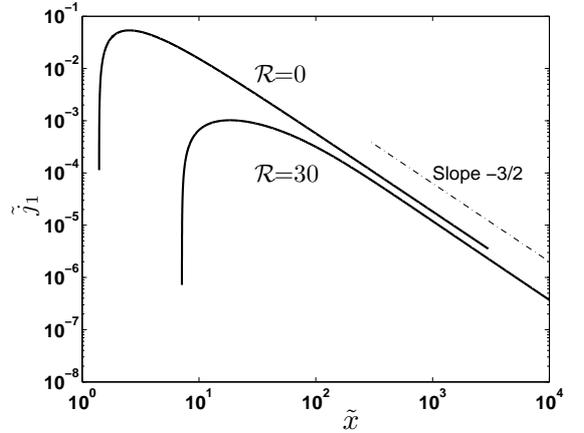}\\(b)
\caption{ First order mass flux term $\tilde{j}_1$. Small (a) and large (b) $\tilde x$ behavior for $\mathcal {R}=0$ and 30. 
Open circles in (a) correspond to the purely kinetic regime, equation (\ref{Solution_flux_only_R}) with $\mathcal {R}=30$.
Only positive values are shown in (b). } \label{j1ll}
\end{figure}
First order terms for curvature $\mathcal K_1$ and mass flux $\tilde{j_1}$ are computed by solving numerically Eq.(\ref{Eq_hydro_nond_first_order_res_T_fin}).  They are displayed in figures  \ref{Curv1} and \ref{j1ll}, respectively.

In the microregion (for $\tilde{x}<\xi$), the curvature is nearly constant, the pressure gradient goes to zero at the contact line, and the hydrodynamic singularity is relieved. The contact line motion and associated evaporation-condensation induce the wedge curvature that is quite large in the contact line vicinity. The contact line curvature grows with $\mathcal {R}$.
The solution for $\mathcal {R}\to\infty$ turns out to be a very good approximation for finite $\mathcal {R} \gg 1$: the solutions for $\mathcal {R}\to\infty$ and $\mathcal {R}=30$ nearly coincide.
Whatever the value of $\mathcal {R}$, the curvature follows the Cox-Voinov asymptotics (\ref{CIR}) for $\tilde{x}\to \infty$.

The mass flux at the contact line is finite too, cf. Fig. \ref{j1ll}.
The mechanism proposed qualitatively by Wayner \cite{Wayner93} to explain contact line motion can be easily visualized. For any finite $\mathcal {R}$,
the mass flux is negative (i.e. condensation for advancing, evaporation for receding) close to the contact line, while its sign changes at the remaining part of the interface.
The overall mass exchange is zero.
From the numerical simulations, $\tilde{j_1}(\tilde {x})\sim\tilde {x}^{-3/2}$ at large $\tilde {x}$ (see Fig. \ref{j1ll}b).

The flux behavior is different for the infinite $\mathcal {R}$ (purely kinetically controlled case).
The flux \eqref{Solution_flux_only_R} remains negative and scales as $\tilde{j_1}(\tilde {x})\sim -\tilde {x}^{-1}$ at large $\tilde {x}$.
The vapor mass conservation (provided by the diffusion equation that does not apply to this case, cf. sec. \ref{conserv}) is not satisfied in the purely kinetically controlled case.
For large but finite $\mathcal {R}$, $\tilde{j_1}(\tilde {x})$ follows the purely kinetic curve (see Fig. \ref{j1ll}a) until some $\tilde {x}$ where a crossover to the purely diffusive regime ($\mathcal {R}=0$) occurs
(see Fig. \ref{j1ll}b), after a sign reversal.
This shows the importance of the diffusion effect that provides the vapor (and therefore liquid) mass conservation during the contact line motion.

\section{Voinov length}  \label{NumSolGen}

\begin{figure}
\centering\includegraphics[width=0.4\textwidth,clip]{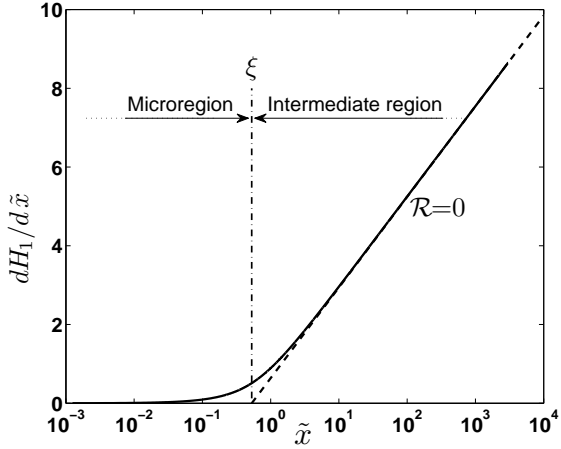}
\caption{Voinov length determination.
  Solid line: first order slope $ dH_1 / d\tilde{x}$;
  dashed line: intermediate region asymptotics \eqref{Cox_Voinovpr};
  vertical dash-dotted line: boundary between micro- and intermediate regions.
} \label{S1diff}
\end{figure}

Figure \ref{S1diff} shows an example of the meniscus slope variation calculated for $\mathcal R=0$.
One can see that at $\tilde{x}\to\infty$ the solution matches the classical asymptotics \eqref{Cox_Voinovpr}.
From the latter, it is evident that the Voinov length $\xi$ corresponds
to the intersection of the $\tilde{x}\to\infty$ asymptote with the $\tilde{x}$ axis.
For a finite $\mathcal R$, the slope variation is similar to Fig. \ref{S1diff}. However, increasing $\mathcal R$ reduces the Voinov length.

The Voinov length can be obtained analytically for the purely kinetic case where the slope is easily deduced from eq. \eqref{Solution_flux_only_R_curv} by integration,
\begin{equation} \label{sloc}
\frac{dH_1}{d\tilde{x}}=-2\sqrt{\tilde{x}\mathcal{R}}K_1\left(\frac{2}
{\sqrt{\tilde{x}\mathcal{R}}}\right)+\mathcal{R}\tilde{x}+\alpha.
\end{equation}
${dH_1}/{d\tilde{x}}$  goes to zero for $\tilde{x}\to 0$ which fixes $\alpha=0$.

A series expansion of equation (\ref{sloc}) at $\tilde{x}\to \infty$ gives
\begin{equation}\label{sloc1}
	\frac{dH_1}{d\tilde{x}}\simeq \ln\left(\frac{\tilde{x}}{\xi}\right) + \frac{\ln \left( e^{3/2} \,  \tilde{x} / \xi\right)}{2 \, e^{2 \gamma-1} \; \tilde{x}/\xi}
\end{equation}
with
\begin{equation}\label{Voinov_length_Ri}
\xi=\frac{e^{2\gamma -1}}{\mathcal{R}} \simeq \frac{1.167}{\mathcal{R}},
\end{equation}
where $\gamma\simeq 0.577216$ is  the Euler-Mascheroni constant.
The first term of the development (\ref{sloc1}) is exactly the asymptotic solution (\ref{Cox_Voinovpr}) in the intermediate region.
The second term is negligible compared to the first for $ \tilde {x} \gg \xi $, which means that for large $\mathcal{R}$, the Kelvin effect is dominant over a distance of the order of $\xi$.

\begin{figure}
\centering\includegraphics[width=0.4\textwidth,clip]{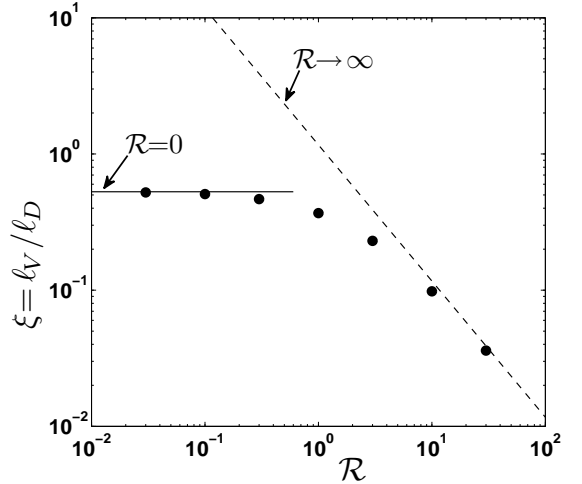}%
  \caption{Dimensionless Voinov length $\xi$ as a function of $\mathcal R$;
circles: numerical result; the solid line is obtained numerically for the purely diffusive regime ($\mathcal {R} = 0$, see Fig. \ref{S1diff}); the dashed line is equation (\ref{Voinov_length_Ri}).} \label{S1}
\end{figure}

The Voinov length is given in Fig. \ref{S1} as a function of ${\mathcal {R}}$.
Two regimes can be observed: for small ${\mathcal {R}}$, when the diffusion dominates, $\xi\sim 1$. For large ${\mathcal {R}}$, $\xi$ decreases as $1/ \mathcal {R}$, as predicted by the analytical equation (\ref{Voinov_length_Ri}).
The crossover $\mathcal {R}$ value is around unity, as expected.
$\xi$  vanishes and relaxation of the contact line singularity is not possible when $\mathcal{R}$ is infinite because the mass flux is brought to zero by the infinite kinetic resistance.

\section{Numerical estimations} \label{usliq}

We have shown that, from the mathematical point of view, the Kelvin effect regularizes the problem.
However this is not enough to conclude on the validity of this approach. It is important to find out by estimation if the continuum theory we developed is valid.
We perform numerical estimations for three common fluids (ethanol, water and glycerol) 
at ambient temperature ($T=298.15$ K) and pressure ($P=1$ bar). The results are gathered in table \ref{ethawat}, for two values of the equilibrium contact angle $\theta $.

In the framework of continuum mechanics approach considered in this paper, two conditions must be fulfilled.
First, when diffusion is the dominant mechanism, the Voinov length $\ell_V$ must be greater than the mean free path in the gas.
Second, the liquid height $\ell_V \theta $ in the microregion, where Kelvin effect acts,  must be larger than the liquid molecule size, of the order of $1$ nm.

The first condition is not restrictive.
Indeed, following standard results from gas kinetic theory, $R_i \sim v_T ^{-1}$ and $D_g \sim v_T \ell_p$,
with $v_T$ the thermal velocity and $\ell _p$ the mean free path of the molecules.
Therefore, $\ell _p / \ell_D \sim \mathcal R$, and the condition $\ell _p \ll \ell_D$ is always fulfilled when diffusion dominates ($\mathcal R \ll 1 $).
In the opposite case ($\mathcal R \gg 1 $), the contact line motion is controlled by the kinetic resistance and diffusion can be disregarded.

\begin{table*}[htb]
  \centering
  \begin{tabular}{lcccccc}
  \hline
  & \multicolumn{2}{c} {Ethanol}   & \multicolumn{2}{c} {Water} & \multicolumn{2}{c} {Glycerol} \\
  \hline
  $\rho$ ($\mathrm{kg.m^{-3}}$)       & \multicolumn{2}{c} {785} & \multicolumn{2}{c} {997} & \multicolumn{2}{c} {1258} \\
  $D_g$ ($\mathrm{mm^{2}.s^{-1}}$)    &  \multicolumn{2}{c} {$12$} & \multicolumn{2}{c} {$26$} & \multicolumn{2}{c} {$8.8$}  \\
  $\mu$ ($\mathrm{mPa.s}$)            & \multicolumn{2}{c} {$1.08$} & \multicolumn{2}{c} {$0.890$} & \multicolumn{2}{c} {$945$}  \\
  $\sigma$ ($\mathrm{mN.m^{-1}}$)     & \multicolumn{2}{c} {$21.9$} & \multicolumn{2}{c} {$71.8$} & \multicolumn{2}{c} {$63.3$}  \\
  $c_{eq}$ ($\mathrm{kg.m^{-3}}$)     & \multicolumn{2}{c} {0.147} & \multicolumn{2}{c} {0.0230} & \multicolumn{2}{c} {$7.43 \times 10^{-7}$}  \\
  $M$ ($\mathrm{g.mol^{-1}}$)                  & \multicolumn{2}{c} {46.07} & \multicolumn{2}{c} {18.02}  & \multicolumn{2}{c} {92.10} \\
  $R_i$  ($\mathrm{s.m^{-1}}$)                 & \multicolumn{2}{c} {0.011} & \multicolumn{2}{c} {0.0068} & \multicolumn{2}{c} {0.015}  \\
  $\theta$ & $1^\circ$ & $5^\circ$    & $1^\circ$ & $5^\circ$ & $1^\circ$ & $5^\circ$ \\
  $J \; (\mathrm{kg.m^{-2}.s^{-1}})$  &  $4.92 \times 10^{-4}$  &              $0.308$    &  $2.49 \times 10^{-3}$  &  $1.55$                 &  $2.61 \times 10^{-6}$  &  $1.63 \times 10^{-3}$ \\
  $C \; (\mathrm{kg.m^{-3}})       $  &  $7.38 \times 10^{-6}$  &  $4.12 \times 10^{-4}$  &  $4.49 \times 10^{-6}$  &  $2.51 \times 10^{-4}$  &  $2.68 \times 10^{-9}$  &  $1.50 \times 10^{-7}$ \\
  $\ell_D$ ($\mathrm{nm}$)            & $180$  & $16$   & $47$   & $4.2$  & $9.0$  & $0.81$ \\
  $\ell_R$ ($\mathrm{nm}$)            & $249$  & $2.0$  & $13$   & $0.10$ & $0.61$ & $0.0049$\\
  $\mathcal R $                       & $0.72$ & $8.1$  & $3.7$  & $42$   & $15$   & $166$\\
  $\ell_V $ ($\mathrm{nm}$)           & $72$   & $1.9$  & $9.5$  & $0.12$ & $0.63$ & $0.0057$\\
  $\ell_V \theta $ ($\mathrm{nm}$)    & $1.3$  & $0.16$ & $0.16$ & $0.01$ & $0.01$ & $0.0005$\\
\end{tabular}
\caption{Physical properties \citep{Marcus,Cussler},
kinetic resistance $R_i$ (using equation (\ref{HertzKnud}) with $f = 2/3$), 
mass flux and concentration scales $J$ and $C$, 
characteristic lengths $\ell_D$ and $\ell_R$,
dimensionless kinetic resistance $\mathcal R $,
Voinov length $\ell_V $
and liquid height in the microregion $\ell_V \theta $,
for ethanol, water and glycerol at ambient conditions ($T=298.15$ K and $P=1$ bar).}\label{ethawat}
\end{table*}

On the contrary, one can see from table \ref{ethawat} that the second condition is not fulfilled, at least for the considered fluids at ambient conditions,
as the height $\theta \ell_V$ is never much larger than 1 nm, and much lower in most cases.
Notice that the lowest values are obtained with glycerol, mainly because of its low volatility
(the Voinov length $\ell_V$ goes to zero for vanishing saturation vapor pressure). 

\section{Conclusion} \label{Conclusion}

A geometrical contact line singularity appearing in the partial wetting regime manifests itself in the hydrodynamic problem and should be relaxed in any theoretical approach. By considering the volatile fluid case, we address in this article a possibility to relax it by the interfacial phase change (evaporation-condensation), which is a mechanism first outlined by Wayner \cite{Wayner93}. We propose a rigorous solution for the case of diffusion-controlled evaporation, when an inert gas is present in the atmosphere. Our work follows recent studies dedicated to the case of a liquid surrounded by its pure vapor \cite{EuLet12,Rednikov13,PRE13movingCL}, where the phase change is controlled by the latent heat effect.

An account of the Kelvin effect is necessary to couple the vapor concentration variation and the liquid meniscus curvature in the contact line vicinity.
It has been found that within a continuum mechanics formulation, based on the lubrication approximation for the liquid dynamics and stationary vapor diffusion in the saturated atmosphere, the contact line singularity is relieved and all the physical quantities (meniscus curvature, mass flux, etc.) become large but finite at the contact line. Since the mass flux is large there, accounting for the interfacial resistance to evaporation/condensation is necessary. During the contact line advancing motion over the dry substrate, condensation occurs at the liquid wedge tip, while exactly the same quantity of liquid is evaporated from the other part of the liquid meniscus so that the liquid mass is conserved. The opposite mass transfer occurs at the receding motion. The obtained wedge slope matches the Cox-Voinov classical solution far from the contact line (in the intermediate asymptotic region). A characteristic (Voinov) length of such a process corresponds to a distance at which the mass transfer occurs. The Voinov length is found as a function of the relative contribution of diffusion and interfacial resistance effects defined by a dimensionless parameter. In a case where the mass transfer is dominated by the interfacial resistance an analytical solution is found.

Numerical estimations show however that for three common fluids (ethanol, water and glycerol) under  the ambient conditions, the Voinov length is very small,
leading to inconsistency of the model with the continuum mechanics (in framework of which it is however developed).
It is not however excluded that the phase change solves the singularity at the molecular scale, within a discrete, e.g. molecular dynamics approach. 

The behavior observed with the contact line singularity relaxation by the diffusion-controlled phase change is qualitatively similar to previous results \citep{EuLet12,Rednikov13,PRE13movingCL} obtained for the pure vapor atmosphere. 
In the latter case, the phase change regularizes the contact line singularity and the approach is consistent with the continuum model for small contact angles \citep{PRE13movingCL}.

\section*{Acknowledgements}
This work has been financially supported by the LabeX LaSIPS (ANR-10-LABX-0040-LaSIPS) managed by the French National Research Agency under the ``Investissements d'avenir'' program (ANR-11-IDEX-0003-02). VN and VJ are grateful to B. Andreotti for fruitful discussions.

\appendix
\section{Governing equation for the vapor concentration field}\label{appA}
The tilde denoting dimensionless quantities is omitted in this appendix.
Here is the derivation of the equation connecting the reduced vapor concentration $c$ at the interface and the mass flux:
\begin{equation}\label{convolution_A}
 c( {x'}, {y'}=0)=-\int_{0}^{\infty}G( x,x') {j}( x)d x,
\end{equation}
where $G$ is the Green function.
The starting point for the formula (\ref{convolution_A}) derivation is Green's second
identity
\begin{multline}\label{Green2}
\int_{\mathcal D} [c(\vec{r})\Delta_r G(\vec{r},\vec{r'})-G(\vec{r},\vec{r'})\Delta c(\vec{r})]d\vec{r}=\\\oint_{\mathcal L} [c(\vec{r})\nabla_r G(\vec{r},\vec{r'})-G(\vec{r},\vec{r'})\nabla c(\vec{r})]\cdot\vec{n} dl_r,\end{multline}
where $\mathcal L$ is the boundary of a domain $\mathcal D$ with the outward unit normal $\vec{n}$. The boundary consists of the $x$ axis $\mathcal{L}_x$ and the line $\mathcal C$, which is a half-circle of infinite radius in the upper half-plane.
The differentiation in all differential operators is assumed hereafter to be performed over the components of the vector $\vec{r}$ rather than those of $\vec{r'}$.

The equation for $c$ is
\begin{equation}\label{eqc}
  \Delta c=0
\end{equation}
with the boundary conditions at $\vec{r}\in\mathcal{L}_x$,
\begin{align}\label{bc}\begin{aligned}
  \frac{\partial c}{\partial y} & =0,\quad x<0 \\
  \frac{\partial c}{\partial y} & =-j(x),\quad x>0
\end{aligned}
\end{align}
and
\begin{equation}\label{cinf}
c(\vec{r})=0
\end{equation}
at $\vec{r}\in\mathcal C$. The corresponding to this problem Green function $G=G(\vec{r},\vec{r'})$ satisfies the equation
\begin{equation}\label{eqG}
  \Delta G(\vec{r},\vec{r'})=\delta(\vec{r}-\vec{r'}).
\end{equation}
Its general solution in 2D is
\begin{equation}\label{Gf}
  G(\vec{r},\vec{r'})=\frac{1}{2\pi}\ln |\vec{r}-\vec{r'}|+H(\vec{r},\vec{r'}),
\end{equation}
where $H$ satisfies the equation $\Delta H=0$ in $\mathcal D$ so it is nonsingular when $\vec{r}=\vec{r'}$. It is determined from the boundary conditions for $G$.  Equation \eqref{Gf} can be easily derived from the divergence theorem  \eqref{dt} applied to $G(|\vec{r}-\vec{r'}|)$ inside a circle of radius $R$ centered at $\vec{r'}$. Equation \eqref{dt} reduces to
\begin{equation*}
 1=2\pi R\frac{dG(R)}{dR},
\end{equation*}
from which one obtains directly \eqref{Gf}.

Let us solve equation \eqref{eqG} with the boundary condition
\begin{equation}
\left.\frac{\partial G}{\partial\vec{n}}\right|_{\vec{r}\in\mathcal{L}_x}=0. \label{Gbc}
\end{equation}
We must find $H$ for the domain $\mathcal D$ which is the upper half plane.
One may use the mirror reflection method. One places another source at the point $\vec{r''}=(x', -y')$
that situates in the lower half plane
(i.e. one solves $\Delta H(\vec{r},\vec{r''})=\delta(\vec{r}-\vec{r''})$), which permits to satisfy
the condition \eqref{Gbc}. One obtains
\begin{multline}\label{Ghalf}
G(x,x',y,y')=\frac{1}{2\pi}(\ln |\vec{r}-\vec{r'}|+\ln |\vec{r}-\vec{r''}|)\\\equiv\frac{1}{4\pi}\ln \{[(x-x')^2+(y-y')^2][(x-x')^2+(y+y')^2]\}.
\end{multline}
Evidently, the condition $\Delta H=0$ is satisfied in the upper half plane because $y'>0$.

The substitution of Eqs. (\ref{eqc}) and (\ref{eqG}) into the lhs of \eqref{Green2} results in the equality
\begin{equation}\label{Green3}
\left\{\begin{array}{cc}
  c(\vec{r'}), &\vec{r'}\in \mathcal D \\
  0,&\mbox{otherwise}
\end{array}\right\}
 =\int_{\mathcal{L}_x} [c(\vec{r})\nabla_r G(\vec{r},\vec{r'})-G(\vec{r},\vec{r'})\nabla c(\vec{r})]\cdot\vec{n} dl_r.
\end{equation}
Note that the contribution of the infinite half-circle $\mathcal C$ to the contour integral is zero.
Indeed, the first term vanishes  because of the condition \eqref{cinf}.
The second term is zero because the flux $\partial c/\partial r$
vanishes at $r\to\infty$ faster than $r^{-1}$, as demonstrated in section \ref{conserv}.

\begin{figure}
   \centering
  \includegraphics[width=0.6\columnwidth,clip]{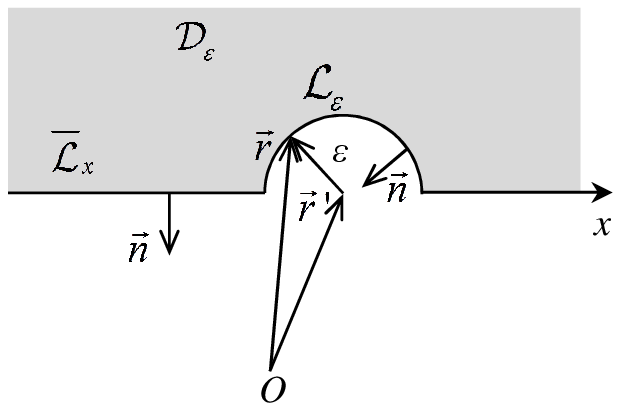}
\caption{A contour sketch.}\label{GContour}
\end{figure}

To obtain \eqref{convolution_A}, one needs to apply \eqref{Green3} at $y'=0$, i.e.,
with $\vec{r'}$ situating exactly at the $\mathcal D$ boundary.
The boundary integral theory suggests that there might be surprises there because of singularity of the kernel $\nabla_r G(\vec{r},\vec{r'})$ when $\vec{r}=\vec{r'}$. To check it, let us replace the contour
$\mathcal{L}_x$ by a contour $\bar {\mathcal{L}}_x\cup \mathcal{L}_\varepsilon$
shown in Fig. \ref{GContour}, where $\mathcal{L}_\varepsilon$ is a half circle of radius $\varepsilon$ centered at $\vec{r'}$. The new domain $\mathcal{D}_\varepsilon$ lies above the contour and becomes $\mathcal{D}$ in
the limit $\varepsilon\to 0$. Let us apply equation \eqref{Green3} to the domain $\mathcal{D}_\varepsilon$. Since $\vec{r'}$ does not belong to it, the lhs is 0.
Let us calculate the integral in rhs over $\mathcal L_\varepsilon$ in the limit $\varepsilon\to 0$. It is evident that
\begin{equation}\label{gradLn}
\nabla_r G(\vec{r},\vec{r'})=\frac{\vec{r}-\vec{r'}}{\pi|\vec{r}-\vec{r'}|^2}.
\end{equation}
It is taken into account here that at the considered boundary $\vec{r'}=\vec{r''}$; the above value is thus double of its counterpart that would be obtained if the free space Green function were used. Note also that the condition
\begin{equation}\label{Gbc1}
\frac{\partial G}{\partial\vec{n}}=0
\end{equation}
holds at $|\vec{r}|\to\infty$.

One can see now that in the limit $\varepsilon\to 0$
\begin{equation}\label{intGradLe}
\int_{\mathcal L_\varepsilon} c(\vec{r})\nabla_r G(\vec{r},\vec{r'})\cdot\vec{n} dl_r=c(\vec{r'})\int_0^\pi\frac{-\varepsilon}{\pi\varepsilon^2}\varepsilon d\phi=-c(\vec{r'}),
\end{equation}
where $\phi$ is the polar angle because $dl_r=\varepsilon d\phi$ and $(\vec{r}-\vec{r'})\cdot\vec{n}=-\varepsilon$.
Note that, if the Green function for the infinite space were used like in the boundary integral theory, the factor $1/2$ would appear near $c$.

It is easy to check that the contribution of the second term of equation \eqref{Green3} (of the integral
over $\mathcal{L}_\varepsilon$) is $O(\varepsilon\ln\varepsilon)\to 0$. When $\varepsilon\to 0$, the contour $\bar {\mathcal{L}}_x$ tends to $\mathcal{L}_x$ and one obtains finally that
the first option of equation \eqref{Green3} is valid also when $\vec{r'}\in \mathcal{L}_x$,
\begin{multline}\label{Green4}
c(\vec{r'})=\int_{\mathcal L_x} [c(\vec{r})\nabla_r G(\vec{r},\vec{r'})-G(\vec{r},\vec{r'})\nabla c(\vec{r})]\cdot\vec{n} dl_r\\=\int_{-\infty}^{\infty}G(x,y=0,x', y'=0)\left.\frac{dc}{dy}\right|_{y=0}dx.\end{multline}
The latter equality is valid because of the condition \eqref{Gbc}. Note that the integral should be taken in Cauchy's Principal Value sense (since it is a limit of the integral over the $\bar {\mathcal{L}}_x$ contour).
By using the conditions \eqref{bc}, equation \eqref{Green4} reduces to \eqref{convolution_A}.
The function
\begin{equation} \label{expGA}
 G(x,y=0,x', y'=0)=\frac{1}{\pi} \ln{| {x}- {x}'|}
 \end{equation}
obtained with the Green function \eqref{Ghalf} can be used in equation \eqref{concentration_at_interface}.

\section{Solution method for the integro-differential equation}\label{appB}

\begin{figure}
\centering \includegraphics[width=0.8\columnwidth]{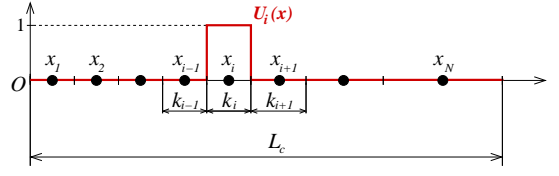}
 \caption{Mesh and interpolation function $U_i(x)$.}\label{MeshInterp}
\end{figure}

The tilde denoting dimensionless quantities is omitted in this appendix.
Equation (\ref{Eq_hydro_nond_first_order_res_T_fin}) is solved numerically. The unknown function $j_1(x)$ is interpolated in the domain $[0,L_c]$, and is assumed to follow a power law in $[L_c , \infty]$, with a cut-off $L_c \gg 1$. The interpolation in the domain $[0, L_c]$ is performed by splitting this interval into $N$ subintervals of length $k_i$ ($i=1$ to $N$),
and by using the interpolation functions $U_i(x)=H(x_i- k_i/2)-H( x_i+ k_i/2)$,
with $H(x)$ the Heaviside function and $x_i$ the center of the $i^{th}$ subinterval (see Fig. \ref{MeshInterp}).
The approximate expression of $j_1(x)$ then reads
\begin{eqnarray}
	j_1(x) = \sum_{i=1 }^N j_i U_i(x) + H(x - L_c) \alpha x ^{-p}.
\label{UnknownFunction}
\end{eqnarray}
The hypothesis about the power law behavior at large $x$ is based on the preliminary numerical studies with $\alpha=0$ and increasingly large $L_c$. They suggested the power law behavior but converged poorly.

One needs to determine $(N+2)$ unknowns: $j_i$ for $i=1\dots N$, $\alpha$ and $p$.
$N$ equations are provided by writing equation (\ref{Eq_hydro_nond_first_order_res_T_fin}) at the nodes $x=x_i$, one more by the continuity of solution at $x=L_c$, and the last by the mass balance (\ref{mconsx}).

The integrals in equation (\ref{Eq_hydro_nond_first_order_res_T_fin}) are expressed using relation \eqref{UnknownFunction},
\begin{multline}\label{DiscInt1}
\int _0 ^\infty \frac{j_1(x')}{x_i-x'}dx' =\\ \sum _{m=1} ^N j_m \int_{x_m-k_m/2} ^{x_m+k_m/2} \frac{dx'}{x_i-x'}
+  \alpha \int_{L_c} ^{\infty} \frac{dx'}{x'^p (x_i-x')}.
\end{multline}
Note that a vanishing for $i=m$ denominator under the first integral in the r.h.s. is not a problem: the integral has a zero Cauchy principal value. Finally, the integral reads
\begin{multline}
\label{DiscInt2}
	\int _0 ^\infty \frac{j_1(x')}{x_i-x'}dx' \\= \sum _{m=1} ^N j_m \ln{\left| \frac{x_m-x_i-k_m/2} {x_m-x_i+k_m/2} \right|}
	-  \alpha \, x_i^{-p} \, \mathrm{Beta}\left(\frac{x_i}{L_c},p,0\right),
\end{multline}
with $\mathrm{Beta}(z,a,b)=\int_0 ^z t^{a-1} (1-t)^{b-1} dt$ the incomplete Beta function.
The second integral of equation (\ref{Eq_hydro_nond_first_order_res_T_fin}) reads
\begin{eqnarray}
\label{DiscInt3}
\int _0 ^{x_i} j_1(x')dx' = \sum _{m=1} ^{i-1} j_m k_m + j_i \frac{k_i}{2}.
\end{eqnarray}
The interfacial resistance term in (\ref{Eq_hydro_nond_first_order_res_T_fin}) is discretized with the first order finite difference scheme
\begin{equation}
\label{DiscInt4}
\left(-\mathcal Rx^3\frac{dj_1}{dx}\right)_i \simeq - {\mathcal R} x_i^3 \frac{j_{i+1}-j_{i}}{x_{i+1}-x_{i}},
\end{equation}
with $x_{N+1}=L_c$ and $j_{N+1}=\alpha L_c^{-p}$. Equation (\ref{Eq_hydro_nond_first_order_res_T_fin}) is discretized using relations (\ref{DiscInt2}-\ref{DiscInt4}), to get $N$ equations
\begin{multline}
\label{AlgEq1}
\frac {x_i^3}{\pi}  \sum _{m=1} ^N {j_m} \ln{\left| \frac{x_m-x_i-k_m/2} {x_m-x_i+k_m/2}\right|}- {\mathcal R} x_i^3 \frac{j_{i+1}-j_{i}}{x_{i+1}-x_{i}}\\
+\sum _{m=1} ^{i-1} j_m k_m + j_i \frac{k_i}{2}=-x_i + \alpha \frac{x_i^{-p+3}}{\pi} \, \mathrm{Beta}\left(\frac{x_i}{L_c},p,0\right).
\end{multline}

The continuity with the power law at large $x$ could be insured by simply writing $(N+1)$-th equation as $j_N = \alpha x_N^{-p}$.
However, to improve the algorithm convergence, we use instead a weaker condition imposed on the last $n_f$ node values. Minimizing the function $S(\alpha)=\sum _{m=N-n_f} ^N (\alpha x_m^{-p}-j_m)^2$ gives the $(N+1)$-th algebraic equation

\begin{eqnarray}
	\label{AlgEq2}
	\sum _{m=N-n_f} ^N j_m x_m^{-p} = \alpha \sum_{m=N-n_f}^N x_m^{-2 p}.
\end{eqnarray}
Finally, the $(N+2)$-th algebraic equation is obtained by discretizing the liquid mass balance \eqref{mconsx} that reduces to $\int _0 ^\infty j_1 (x) dx=0$,
\begin{eqnarray}
	\label{AlgEq3}
	\sum _{m=1} ^N j_m k_m = \alpha \frac{L_{c}^{1-p}}{1-p}.
\end{eqnarray}
The latter expression assumes $p\neq 1$.

\begin{figure}
 \centering \includegraphics[width=0.8\columnwidth]{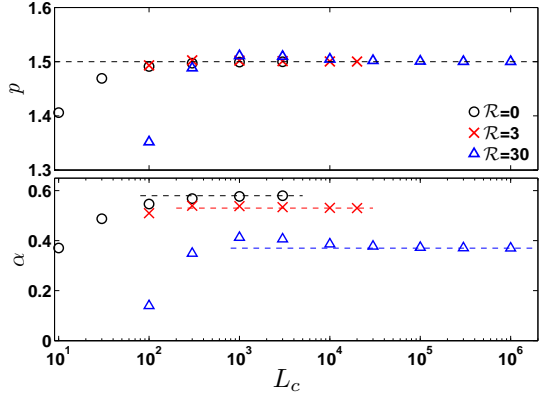}
 \caption{Convergence test of the numerical algorithm. Exponent $p$ (top) and prefactor $\alpha $ (bottom) as functions of the cut-off $L_c$ for three values of $\mathcal R$.
 Dashed lines represent the asymptotes reached at large $L_c$.}\label{conv}
\end{figure}

To test the validity of the numerical approach, the convergence of the prefactor $\alpha $ and the exponent $p$ with respect to the cut-off $L_c$ have been checked, see Fig. \ref{conv}.
Convergence is attained for $L_c\gtrsim 300$ for $\mathcal R=0$ and $3$, and $L_c\gtrsim 10^4$ for $\mathcal R=30$. Fig. \ref{conv} shows $p=1.5$ which corresponds to the power law behavior at large $x$ in Fig. \ref{j1ll}b.

\section{Numerical procedure to get $\mathcal K_1$ from $j_1$}\label{appC}

The tilde denoting dimensionless quantities is omitted in this appendix.
Computing the curvature $\mathcal K_1(x_i)$ from equation (\ref{curvature_convolution})
requires numerical evaluation of an integral (\ref{expGA}) which involves the Green function $G(x_i,x')$. Using the discretization of $j_1$ given by equation (\ref{UnknownFunction}), one gets
\begin{multline}
\label{IntegGJA}
\int_0 ^{\infty} G(x_i,x') j_1(x') dx' \simeq \sum _{m=1} ^{N} j_{m} \int_{x_m - k_m/2} ^{x_m + k_m/2} G(x_i,x') dx'\\
+ \alpha \int_{L_c} ^{\infty}  \frac{G(x_i,x')} {x'^{p}} dx'.
\end{multline}
A Cauchy principal value can be assigned to the first integral of the r.h.s. of equation (\ref{IntegGJA}) when $m = i$. Both integrals can be computed analytically,
\begin{align}
\label{AntiDerGA1}
 \int_{x_m - k_m/2}^{x_m + k_m/2} G(x_i,x')  dx' = \frac{1}{\pi} \Biggl[ \left(x_m-x_i+\frac{k_m}{2}\right) \ln \left | x_m-x_i+\frac{k_m}{2} \right | \nonumber & \\
 -\left(x_m-x_i-\frac{k_m}{2}\right) \ln \left | x_m-x_i-\frac{k_m}{2} \right | -k_m \Biggr] , &\\
\label{AntiDerGA2}
 \int_{L_c} ^{\infty}  \frac{G(x_i,x')} {x'^{p}} dx' = \frac{1}{\pi (p-1)} \Biggl[L_c^{1-p} \ln (L_c-x_i) \nonumber & \\ +x_i ^{1-p}\, \mathrm{Beta}\left(\frac{x_i}{L_c},p-1,0\right) \Biggr]. &
\end{align}

%\bibliographystyle{elsarticle-num-names}
%\bibliography{ContactTransf,Books,DiffEvap}

\end{document}